\documentclass[aps,prl,a4paper,twocolumn,showpacs]{revtex4}
\usepackage{epsfig}

\begin{document}

\title{First Principle Electronic Model for High-Temperature
Superconductivity}

\author{V.I.~Anisimov}
\author{M.A.~Korotin}
\author{I.A.~Nekrasov}
\author{Z.V.~Pchelkina}
\affiliation{
Institute of Metal Physics, Ekaterinburg GSP-170, Russia}
\author{S.~Sorella}
\affiliation{
Istituto Nazionale per la Fisica della Materia, and SISSA,
I-34014 Trieste, Italy}

\date{06 March 2002}
\begin{abstract}

Using the structural data of the $La_2 Cu O_4$ compound both in the LTT
phase and in the isotropic phase we have derived an effective $t-J$
model with hoppings $t$ and superexchange interactions $J$ extended up
to fourth and second neareast neighbors respectively.  By numerically
studying this hamiltonian we have then reproduced the main experimental
features of this HTc compound:  d-wave superconductivity is stabilized
at small but finite doping $\delta > 6\%$ away from the
antiferromagnetic region and some evidence of dynamical stripes is found
at commensurate filling $1/8$.

\end{abstract}

\maketitle

The microscopic mechanism of high-temperature superconductivity (HTc) is
probably the most important, but still open problem in condensed matter
physics.  After the recent discovery of $Mg B_2$~\cite{mgb2} -- a phonon
like superconductor with $Tc\simeq 39K$, and the fullerene
superconductivity temperature~\cite{battlog} has been substantially
enhanced in a compound where certainly the electron-phonon coupling
cannot be neglected, there have been increasing expectations~\cite{shen}
that the conventional type of electron-phonon BCS superconductivity may
explain all HTc materials~\cite{shen}.  On the other hand, from the
theoretical point of view, a wide range of numerical techniques have
recently led~\cite{recpwa} to rather consistent evidence that strongly
correlated models, such as Hubbard like~\cite{prush}, and $t-J$
models~\cite{dagotto,white} may show d-wave superconductivity at low
temperature, evidence that is rather remarkable because no
electron-phonon or explicit attractions are included in these models.

In order to contribute to the solution of these controversial aspects
between theories and experiments and to understand the role of the
strong electron correlation, in the present work we try to reduce the
distance between the physics of rather abstract model hamiltonians and a
consistent \textit{ab-initio} derivation of experimental properties of HTc
material.  We consider the most popular compound for HTc
superconductivity -- $La_2 Cu O_4$.  We derive, and then study by
Quantum Monte Carlo methods, an effective model in which the
electron-phonon interaction is neglected.  Our purpose is to understand
whether the physics of an effective model hamiltonian may explain
properties of real materials that are not understood with conventional
methods of band theory and/or BCS mechanism of superconductivity.  To
our knowledge our work represents the first attempt to explain the
physics of HTc superconductors, starting from first principle
calculations.

The $La_2 Cu O_4$ compound is known to be superconductor in a range of
doping between $6\%$ and $~30\%$, and by interstitial substitution of Nd
(in place of La) at commensurate filling $1/8$, static incommensurate
peaks in the magnetic structure factor were observed by neutron
scattering experiments~\cite{tranquada} and, correspondingly, the
superconducting transition temperature was strongly suppressed.  The
Nd-substitution enhances the spatial anisotropy and is believed to favor
the formation of stripes:  one dimensional hole-rich patterns in the CuO
planes, separating antiferromagnetic (AF) domains with opposite
direction of the AF order parameter, namely the next nearest neighbor
Cu-spins across the stripe are antiparallel.  This feature easily
explains the mentioned neutron scattering incommensurate peaks, within
the assumption that the stripes are half filled (half an hole per Cu
site density along the stripe).  In fact this implies that, at a given
hole doping $\delta$, the incommensurate peak in the magnetic structure
factor $S(q)$ shifts by $2 \pi \delta$ from the AF wavevector $(\pi,\pi)
\to (\pi,\pi-2 \pi \delta)$ (for equally spaced stripes parallel to the
x-axis direction), correspondingly the charge structure factor $N(q)$
shows up an incommensurate peak close to the $\Gamma$ point for $q=(0,4
\pi \delta)$~\cite{tranquada}.

We consider the extended $t-J$ one-band model on a finite lattice with
$N$ sites:  $${\cal H} = \sum_{ R,\mu } J_{\mu} {\bf S}_R \cdot {\bf
S}_{R+\tau_\mu} - \sum_{ R,\mu, \sigma} t_{\mu} \left( {\tilde
c}^{\dag}_{R,\sigma} {\tilde c}_{R+\tau_\mu,\sigma} +H.c.  \right ) $$
where ${\tilde c}^{\dag}_{i,\sigma}=c^{\dag}_{i,\sigma} \left ( 1-
n_{i,\bar \sigma} \right )$, and ${\bf S}_i$ is the electron spin
operator on site $i$, whereas the sum run over the lattice sites $R$ and
corresponding neighbors $R+\tau_\mu$, determined by vectors $\tau_\mu$
shown in Table I.  Previous work have shown the key relevance of
hoppings extended beyond nearest neighbors to reproduce the
phenomenology of the Cuprates~\cite{naza}.  In this table we
microscopically derive the values of the effective parameters $t_\mu$
and $J_\mu$ determined from the results of electronic structure
calculations by standard LDA (TBLMTO~\cite{lmto}) and LDA+U
method~\cite{lda+u}, for the anisotropic case (with Nd substitution) and
the high-symmetry tetragonal case, using known structural
data~\cite{LTT-structure}.  The effective hopping parameters $t_\mu$ were
calculated by the standard least-square fit procedure to the bands
obtained in LMTO calculations.  The effective exchange parameters
$J_\mu$ were calculated using the formula derived by
A.~Lichtenstein~\cite{lichtexchange,anis}, where the second derivative of
the total energy as a function of the value of the angle between spin
directions on two atoms is calculated via eigenvalues and eigenfunctions
obtained in electronic structure calculations.  Those electronic
structure calculations were done using LDA+U method which allows to
obtain antiferromagnetic insulating ground state for the undoped
cuprate.  By contrast, the standard LDA method leads to a nonmagnetic
metallic ground state~\cite{lda+u}, which is obviously inconsistent with
experiments.  The Coulomb parameters U and J used in LDA+U calculations
were obtained in constrained LDA calculation~\cite{superlsda} (U=11~eV,
J=1~eV).

\begin{table}
\caption {\label{table} Values of the effective parameters in eV.}
\begin{tabular}{ccccc}
\hline
  & Anisotropic &  & Isotropic &    \\
\hline
$\tau$ & $J_\mu $  & $t_\mu$ & $J_\mu$ & $t_\mu$ \\
\hline
 (1,0) &  0.105 &  0.425   & 0.109  &   0.486  \\
 (0,1) &  0.111 &  0.466   & 0.109  &   0.486   \\
 (1,1) &  0.016 &  0.014   & 0.016  &  -0.086   \\
 (2,0) &  0       & 0.036  &  0     &  -0.006   \\
 (0,2) &  0       & -0.064 &  0     &  -0.006    \\
 (2,1) &  0       & -0.001 &  0     &    0       \\
 (1,2) &  0       & 0.046  &  0     &    0       \\
\hline
\end{tabular}
\end{table}

\begin{figure} 
\epsfxsize=85mm 
\centerline{\epsffile{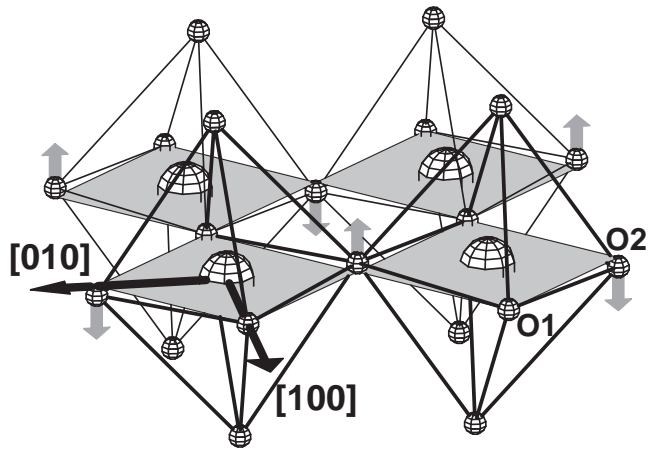}} 
\caption{\label{fig1} Structural data for the
$HTc$ materials considered (see text).  The displacements of the ions
in the anisotropic LTT phase is indicated by the arrows.}  
\end{figure}

The Nd-substitution for La in $La_2 Cu O_4$ results in anisotropic Low
Temperature Tetragonal (LTT) structure~\cite{LTT-structure}
(Fig.\ref{fig1}), where the $CuO_6$ tetrahedra are tilted via
rotation around [1,0] axis (and [0,1] axis in the next plane, so that
the total crystal structure preserves tetragonal symmetry).  As a result
of this tilting there are two kinds of oxygen atoms, O1 which form 180
degree Cu-O-Cu bonds in [1,0] direction and are in the Cu atoms plane,
and O2 atoms, which are shifted out of the plane and form Cu-O-Cu bonds
in [0,1] direction with the decreased bond angle.  Remarkably the
strongest nearest neighbor hopping is in the direction [0,1] where the
oxygen atoms are out of the Cu-Cu plane (see Fig.~\ref{fig1}).  This
seems counterintuitive, because for this direction the Cu-O-Cu bond
angle is less than 180 degree and hence the effective d-d hopping must
be decreased comparing with the straight Cu-O-Cu bond in [1,0]
direction.  Solution of this puzzle is the difference in electrostatic
potential on O1 and O2 crystallographic positions which results in the
energy of 2p-orbitals of out-of-plane O2 atoms being 1.05~eV higher than
for the in-plane O1 atom.  As the effective d-d hopping is proportional
to $t_{pd}^2/\Delta_{pd}$, where $t_{pd}$ is effective p-d hopping
parameter and $\Delta_{pd}$ is energy separation between Cu-3d and O-2p
orbitals, then $\Delta_{pd}$ is significantly smaller for out-of-plane
O2 oxygen resulting in a larger value of the effective d-d hopping
parameter.  Another effect of anisotropic crystal structure distortion
is the long-range nature of model hamiltonian hopping parameters.  For
the isotropic case only two shells of nearest neighbors were needed and
the value of $t_\mu$ for the third neighbor is negligible.  Instead for
the anisotropic case the hopping parameters have to extend at least up
to fourth neighbors for a satisfactory fit of the LMTO bands.  We must
note also, that the usual scaling of the effective exchange parameters
$J_\mu$ as $t_\mu^2$ is not valid here.

In order to study this model hamiltonian we use the recent QMC methods,
that have been shown to work very well for the standard $t-J$
model~\cite{sorella}, allowing to obtain very accurate values of
energies and correlation functions for a wide range of $J/t$ values.
The initial variational wavefunction is chosen to be the most general
Jastrow-BCS real wavefunction $\psi_G= \hat J |BCS>$ projected onto the
subspace with no-doubly occupied sites and fixed number of particles.
After the latter projection the wavefunction is totally symmetric, and a
large number of variational parameters can be optimized consistently
with the symmetries of the model.  This is the first step $p=0$ of our
procedure (SR)~\cite{sorella,recent} (e.g.  $46$ variational parameters
for the $8\times 8$ anisotropic case), which is in principle convergent for
large number $p$ of iterations.  The scheme is based on the fast
convergence properties of the Lanczos technique, that allows to improve
remarkably the accuracy of the best variational guess with few
iterations.  Starting from the $p-$ Lanczos step wavefunction it is also
possible to further improve the variational energy and especially the
quality of the correlation functions, by a further correction scheme,
such as the fixed node (FN)~\cite{ceperley} or the more accurate
stochastic reconfiguration method (SR)~\cite{recent}.  A comparison of
the quality of our approximation, as compared with the conventional
Fixed node technique~\cite{ceperley} is shown in Tab.~\ref{table2}.

\begin{table}
\caption{\label{table2} Variational energy (eV) per Cu site for the 
anisotropic $t-J$ (Tab.~\ref{table}) model for the best variational FN  
and SR ($p=2$) techniques applied to the $p=1$ Lanczos wavefunction. 
The $\sigma=0$ are estimates of the exact zero variance energies 
\protect\cite{sorella}. Error bars are in brackets. } 
\begin{tabular}{ccccc}
\hline
 \# Holes  & VMC  & FN  &  SR  & $\sigma=0$ \\
\hline
0  & -0.06446(1)   &  -0.06614(4) & -0.06637(4) & -0.06665(17) \\
4  & -0.12717(3)   &  -0.1368(1)  & -0.1379(1) & -0.1402(5) \\
6  & -0.15623(4)   &  -0.1679(1)  & -0.1701(2) & -0.1744(14)  \\
8  & -0.18336(4)   &  -0.1972(1)  & -0.1988(3) & -0.2025(7)  \\
10 & -0.20859(4)   &  -0.2237(1)  & -0.2265(3) & -0.2316(6)  \\
16 & -0.27446(6)   &  -0.2929(1)  & -0.2956(4) & -0.3053(11)  \\
26 & -0.35030(6)   &  -0.3684(1)  & -0.3710(4)  & -0.3780(12)  \\
\hline
\end{tabular}
\end{table}

\begin{figure} 
\epsfxsize=85mm 
\centerline{\epsffile{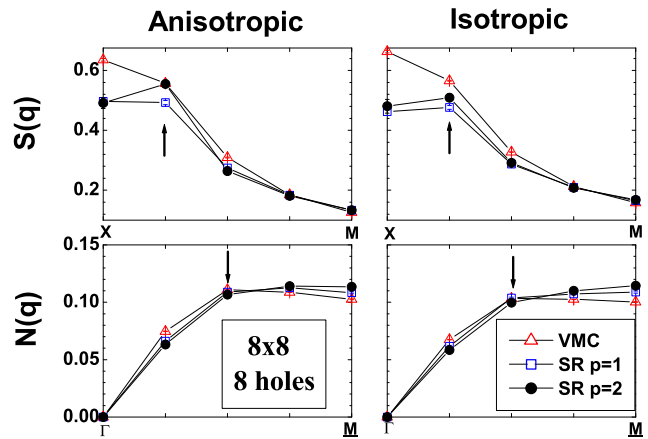}} 
\caption{\label{fig2}
Spin and charge structure factors for the cases of  Tab.~\ref{table} 
vs. improved  accuracy in energy: VMC, SR p=1, SR p=2, respectively.   
$\Gamma=(0,0)$. $X=(\pi,\pi)$, $M=(\pi,0)$, $\underbar M =(0,\pi)$ and the 
arrows indicate the Tranquada's wavevectors (see text).}
\end{figure}

We concentrate our study in the low-doping region and especially at the
important filling $\delta=1/8$ where static stripes were observed.  As
shown in Fig.~\ref{fig2} there is a clear evidence, especially in
the anisotropic case, of an incommensurate peak that was not present at
the variational level, and is sharpening as the accuracy of the
calculation approaches the low energy limit.  Correspondingly the charge
structure factor $N(q)$ does not seem to be enhanced in this limit, but
only a cusp singularity at the Tranquada's wavevector appears consistent
with our data.  The strongest anisotropy appears in the nearest neighbor
hoppings (see Tab.~1) and, correspondingly the peak in the magnetic
structure factor, is shifted by $(0,2 \pi \delta)$, consistent with the
Tranquada's predictions.  Within the ''stripe picture'' this is
compatible with stripes along the $[1,0]$ axis, a direction where the
nearest neighbor hopping is much less favored compared with the other
direction.  Thus, in order to optimize the kinetic energy in the
y-direction, charge-fluctuations perpendicular to the stripes are
required.  These fluctuations suppress any static response in the charge
structure factor, but leave a sizable effect in the spin provided the
spins across the hole are antiparallel~\cite{martins}, as we have
verified in this case.  In the following we name this feature the
''dynamical stripe'', a genuine effect of strong correlation.

As shown in Fig.~\ref{fig2}, the anisotropy clearly enhances the
dynamical stripe fluctuations, since, in the anisotropic case, a more
resolved incommensurate peak for $S(q)$ is found.  Some small effect is
seen also in the isotropic case, consistent with the DMRG
findings~\cite{white} that $t^\prime /t <0$ may stabilize stripes.
Probably at very low energy (larger $p$) evidence of true static stripes
can also be found at this particular filling.  For all other dopings,
though incommensurate magnetic peaks are still present, they are much
less sharp, and the value of $S(q)$ at the maximum is much below the
corresponding $p=0$ variational value.

Regarding superconductivity, we have studied the pairing correlation
functions $\Delta_{i}^{\mu,\nu}(r)$=$\langle {\cal S}_{i+r,\mu} {\cal
S}^\dag_{i,\nu} \rangle.$ Here ${\cal S}^\dag_{i,\mu}$= ${\tilde
c}^\dag_{i,\uparrow} {\tilde c}^\dag_{i+\mu,\downarrow} - {\tilde
c}^\dag_{i,\downarrow} {\tilde c}^\dag_{i+\mu,\uparrow}$ creates an
electron singlet pair in the neighboring sites $(i,i+\mu)$.
Off-diagonal long-range order is implied if $P_d=2 \lim_{r \to \infty}
\sqrt{|\Delta_{i}^{\mu,\nu}(r)}|$ remains finite in the thermodynamic
limit.

\begin{figure} 
\epsfxsize=85mm 
\centerline{\epsffile{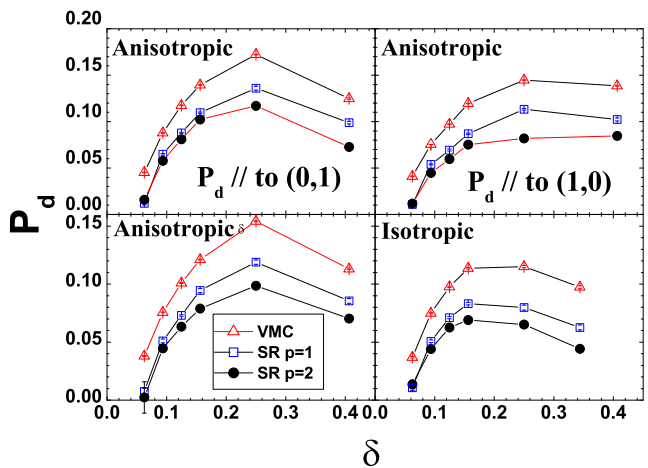}} 
\caption{\label{fig3} $P_d$  for various 
dopings and improved  accuracy in energy, VMC, SR p=1, and SR p=2, 
respectively, for the cases  of Tab.~\ref{table}. 
Lower panels:  the value of $P_d$ was obtained by $d-$wave-averaging 
the pairing  correlations at the maximum distance 
$\protect\simeq 4 \protect\sqrt{2}$.
Upper ones: $P_d$ refers to the maximum  distance ($=4$)  
along the $x,y$  directions.}
\end{figure}

\begin{figure} 
\epsfxsize=85mm 
\centerline{\epsffile{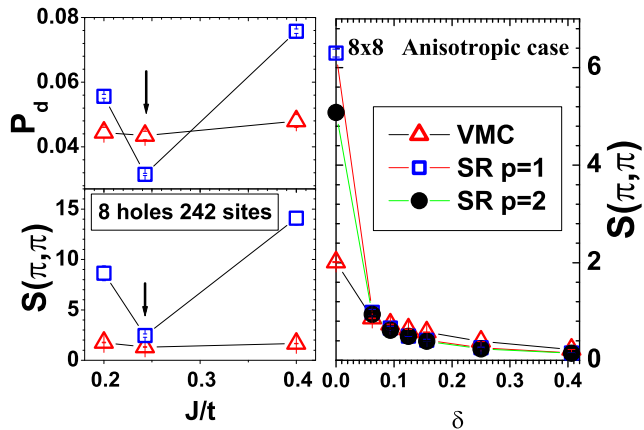}} 
\caption{\label{fig4}  Right panel: $S(\pi,\pi)$, as a function of doping 
$\delta$ for the $8 \times 8$ anisotropic (Tab.~\ref{table}) $t-J$  model. 
Left panels: $S(\pi,\pi)$ and $P_d$ for larger size for  the standard 
$t-J$ model (including  the nearest neighbor density-density interaction, 
which however is irrelevant at small doping) and  the anisotropic  case 
indicated by the arrow. In the latter  case the $J/t$ value is defined 
by the average nearest neighbor ratio.}
\end{figure}

For the $8\times 8$ system, that we have studied for several dopings
(see Fig.~\ref{fig3} lower panels) size effects are acceptable, at
least close to optimal doping  as shown in \cite{dagotto} for the $t-J$
model.  The existence of a finite $P_d$ and correspondingly the absence
of antiferromagnetic long range order is an experimental fact for
$\delta > 6 \%$.  Both features are remarkably well reproduced by our
calculations (see right Fig.~\ref{fig4}).  In particular, at half
filling, the magnetic structure factor dramatically increases with
respect to the spin liquid variational wavefunction $\hat J
|BCS\rangle$, confirming the existence of AF long range order, as widely
accepted for the weakly frustrated Heisenberg model.  Conversely, as
soon as the doping is tinily increased, the AF magnetic structure factor
remains very close to the spin liquid variational reference, clearly
indicating absence of AF long range order.  This is the main result of
our paper and is due to the long range couplings of our effective model,
since at small doping the $t-J$ model without long range couplings is
expected to be both antiferromagnetic and superconductor~\cite{dagotto}.
In order to understand this effect, we have extended the calculation to
larger size $N=242$ at small doping both for the standard $t-J$ model at
$J/t=0.2$ and $J/t=0.4$ and our extended $t-J$ model, in the anisotropic
case.  As it is clearly seen in the left panels of Fig.~\ref{fig4},
the role of the long range hoppings is crucial to destroy both
antiferromagnetism and superconductivity.

Another effect is that, within our model, the anisotropy does not
suppress the superconducting order parameter $P_d$, (see lower
Fig.~\ref{fig3}).  This is rather in contrast with experiments, where
a drastic change of $T_c$ was found in the anisotropic case.  However
in this case the pairing correlations are also anisotropic (see upper
Fig.\ref{fig3}) and $P_d$ vs doping, computed at the maximum $[1,0]$
or $[0,1]$ distance appears to be much suppressed in the direction of
the stripe (notice that this direction is alternatively x or y in
neighboring CuO planes).  It is possible that by including the
electron-phonon coupling one can obtain further agreement with
experiments.  In fact we expect that, whenever dynamical stripes are
clearly formed, the electron-phonon coupling can considerably enhance
the anisotropy and lead to the formation of true static stripes, with
negligible $P_d$.

In conclusion we can reproduce many aspects of the low-doping phase
diagram of HTc compounds by neglecting completely the electron phonon
coupling, within an \textit{ab-initio} derivation of the effective
superexchange and long-range hopping couplings for an extended $t-J$
model.  Several interesting features comes out from our calculation that
deserve experimental confirmation:  the finite critical doping
$\delta_c$ required to stabilize superconductivity, appears a band
structure effect.  In principle antiferromagnetism and superconductivity
may coexist at small enough doping, provided the long range hoppings are
suppressed.  The anisotropy and long range hoppings appear to be
compatible with a sizable superconducting order parameter, as long as
the stripes remain dynamical~\cite{andersen}.  Static stripes are not
clearly stabilized in a model in which the electron-phonon interaction
is disregarded.

This work was partially supported by Russian foundation for Basic
Research grant RFFI-01-02-17063 and by MIUR-COFIN2001.  We acknowledge
T.M.~Rice for stimulating the present scientific collaboration and 
E.~Dagotto for useful discussions.

\end{document}